\documentclass[twocolumn,preprintnumbers,amsmath,amssymbm,prd]{revtex4}
\usepackage{epsfig}
\usepackage{graphicx}

\begin{document}
%\title{Remark on the universality of the quasinormal spectrum of near-extremal Kerr-Newman black holes}
\title{A short remark (with a long title) on the universality of the quasinormal spectrum of near-extremal Kerr-Newman black holes}
\author{Shahar Hod}
\address{The Ruppin Academic Center, Emeq Hefer 40250, Israel}
\address{ }
\address{The Hadassah Institute, Jerusalem 91010, Israel}
\date{\today}

\begin{abstract}
\ \ \ In a recent paper (arXiv:1410.0694) Zilh\~ao, Cardoso,
Herdeiro, Lehner, and Sperhake have studied the nonlinear stability
of Kerr-Newman black holes. We show that their numerical results for
the time evolutions of the spacetime deformations of near-extremal
Kerr-Newman black holes are described extremely well by a {\it
universal} formula for the quasinormal resonances of the black
holes. This formula is expressed in terms of the black-hole physical
parameters: the horizon angular velocity $\Omega_{\text{H}}$ and the
Bekenstein-Hawking temperature $T_{\text{BH}}$.
\end{abstract}
\bigskip
\maketitle

%]

\section{Introduction}

It is well known that the fundamental quasinormal resonances of
near-extremal black holes are described by an extremely simple and
compact formula \cite{Hod1,Hod2}:
\begin{equation}\label{Eq1}
\omega_n=m\Omega_{\text{H}}-i2\pi T_{\text{BH}}(n+{1\over
2}-i\delta)\ \ \ ; \ \ \ n=0,1,2,...\  .
\end{equation}
Here \cite{Noteunit,Notepa}
\begin{equation}\label{Eq2}
\Omega_{\text{H}}={{a}\over{r^2_++a^2}}\ \ \ {\text{and}} \ \ \
T_{\text{BH}}={{r_+-r_-}\over{4\pi(r^2_++a^2)}}
\end{equation}
are respectively the angular velocity of the black-hole horizon and
the Bekenstein-Hawking temperature of the black hole, and $\delta$
is the angular-eigenvalue of the angular Teukolsky equation
\cite{Delta,Teuk}. The parameter $m$ in (\ref{Eq1}) is the azimuthal
harmonic index of the perturbation mode.

It should be emphasized that the formula (\ref{Eq1}) was derived
analytically \cite{Hod1,Hod2} for scalar, electromagnetic, and
gravitational perturbations of rotating Kerr black holes. As for the
charged and rotating Kerr-Newman black holes, this formula was
formally derived {\it only} for scalar perturbation fields. The
restriction to scalar fields in the case of Kerr-Newman black holes
is a direct consequence of the fact that all attempts to decouple
the gravitational and electromagnetic perturbations of generic
Kerr-Newman black holes have failed thus far \cite{Chan}.

Most recently, Zilh\~ao, Cardoso, Herdeiro, Lehner, and Sperhake
\cite{ZCHLS} have studied the nonlinear stability of Kerr-Newman
black holes using fully {\it nonlinear} numerical simulations of the
coupled Einstein-Maxwell equations. The numerical results presented
in \cite{ZCHLS} indicate that, at late times, the spacetime
deformations of the Kerr-Newman black hole are characterized by
damped oscillations (see, in particular, Fig. 4 of \cite{ZCHLS}).

One of the most intriguing results presented in \cite{ZCHLS} is a
numerical evidence for a universal behavior of these late-time
spacetime oscillations (quasinormal resonances). In particular, Fig.
4 of \cite{ZCHLS} presents four time evolutions of quadrupolar
spacetime deformations for four different values of the black-hole
parameters $(a,Q)$, all share the same value of the ratio
$a/a_{\text{max}}=0.99$, where $a_{\text{max}}=\sqrt{M^2-Q^2}$.

Three of these time evolutions, which correspond to the following
kerr-Newman parameters
\begin{equation}\label{Eq3}
(a/M,Q/M)=(0.907,0.4)\ , \ (0.944,0.3)\ , \ (0.99,0)\  ,
\end{equation}
overlap almost perfectly \cite{ZCHLS}. However, as indicated in
\cite{ZCHLS}, the curve describing the time evolution of the fourth
perturbed black hole, which is characterized by
\begin{equation}\label{Eq4}
(a/M,Q/M)=(0.594,0.8)\  ,
\end{equation}
does not coincide with the previous three curves (time evolutions)
\cite{ZCHLS}.

Based on the (almost) perfect agreement between the time evolutions
of the three perturbed Kerr-Newman black holes (\ref{Eq3}), it was
suggested in \cite{ZCHLS} that the quadrupolar quasinormal
resonances of Kerr-Newman black holes with large $a/Q$ ratios are
characterized by a universal behavior of the form
\begin{equation}\label{Eq5}
\omega=\omega(a/\sqrt{M^2-Q^2})\  .
\end{equation}
That is, it was suggested \cite{ZCHLS} that the quasinormal
resonances of Kerr-Newman black holes depend solely (and
universally) on the ratio $a/a_{\text{max}}$.

As discussed above, the almost identical temporal evolutions of the
three perturbed black holes (\ref{Eq3}) are in agreement with the
suggested universality (\ref{Eq5}). However, the time evolution of
the fourth perturbed black hole (\ref{Eq4}), which does {\it not}
coincide with the previous three time evolutions, raises doubts on
the general validity of the suggested universal relation
(\ref{Eq5}).

As we shall now show, {\it all} four temporal evolutions displayed
in Fig. 4 of \cite{ZCHLS} [including the time evolution of the
fourth black hole (\ref{Eq4})!] are perfectly consistent with the
formula (\ref{Eq1}). We first point out that the three black holes
(\ref{Eq3}) are characterized by almost the {\it same} value of the
horizon angular velocity $\Omega_{\text{H}}$ (to an accuracy of
better than $0.5\%$) [see Eqs. (\ref{Eq2}) and (\ref{Eq3})]
\begin{equation}\label{Eq6}
\Omega_{\text{H}}\simeq 0.432M^{-1}\  ,
\end{equation}
whereas the fourth black hole (\ref{Eq4}) is characterized by a {\it
different} value of $\Omega_{\text{H}}$
\begin{equation}\label{Eq7}
\Omega_{\text{H}}\simeq 0.388M^{-1}\  .
\end{equation}

It is important to emphasize that time-dependent spacetime
deformations, as the ones displayed in Fig. 4 of \cite{ZCHLS},
consist of a {\it sum} of damped oscillations. Since, for
near-extremal black holes, these oscillations (quasinormal
resonances) share the {\it same} value of $\Re\omega$ [as suggested
by (\ref{Eq1})], one can accurately determine that value of
$\Re\omega$ directly from the time evolutions presented in Fig. 4 of
\cite{ZCHLS} \cite{Noteim}.

The quadrupolar ($l=m=2$) spacetime oscillations of the three
Kerr-Newman black holes (\ref{Eq3}) displayed in Fig. 4 of
\cite{ZCHLS} are characterized by the (numerically computed)
time-period $T_{\text{num}}\simeq 7.38M$. For comparison, the
time-period $T_{\text{ana}}=2\pi/\Re\omega=2\pi/m\Omega_{\text{H}}$
predicted by the analytical formula (\ref{Eq1}) is
$T_{\text{ana}}\simeq 7.27M$ [see Eq. (\ref{Eq6})]. One therefore
finds a fairly good agreement (to better than $2\%$),
\begin{equation}\label{Eq8}
{{T_{\text{num}}}\over{T{\text{ana}}}}\simeq 1.015\ ,
\end{equation}
between the {\it analytical} formula (\ref{Eq1}) and the {\it
numerical} results of \cite{ZCHLS}.

The quadrupolar spacetime oscillations of the fourth Kerr-Newman
black hole (\ref{Eq4}) displayed in Fig. 4 of \cite{ZCHLS} are
characterized by the (numerically computed) time-period
$T_{\text{num}}\simeq 8.11M$. For comparison, the time-period
$T_{\text{ana}}=2\pi/\Re\omega=2\pi/m\Omega_{\text{H}}$ predicted by
the analytical formula (\ref{Eq1}) is $T_{\text{ana}}\simeq 8.09M$
[see Eq. (\ref{Eq7})]. One therefore finds a remarkably good
agreement (to better than $0.3\%$),
\begin{equation}\label{Eq9}
{{T_{\text{num}}}\over{T{\text{ana}}}}\simeq 1.003\ ,
\end{equation}
between the {\it analytical} formula (\ref{Eq1}) and the {\it
numerical} results of \cite{ZCHLS}.

In summary, in this short remark we have shown that the nonlinear
(numerical) time evolutions of the near-extremal Kerr-Newman
spacetime deformations presented in \cite{ZCHLS} are described
extremely well by the formula (\ref{Eq1}).

It is worth emphasizing again that the formula (\ref{Eq1}) was
formally derived \cite{Hod1,Hod2} for scalar, electromagnetic, and
gravitational perturbations of near-extremal rotating Kerr black
holes, as well as for scalar perturbations of near-extremal
Kerr-Newman black holes. Formally, no analytical proof exists for
its validity in the case of coupled gravitational-electromagnetic
perturbations of Kerr-Newman black holes.

However, the results presented here and in \cite{ZCHLS} strongly
suggest that the formula (\ref{Eq1}) is actually of general
validity. In particular, we have shown that it accurately describes
the coupled gravitational-electromagnetic quasinormal resonances of
charged and rotating near-extremal Kerr-Newman black holes.

\bigskip
\noindent
{\bf ACKNOWLEDGMENTS}
%\bigskip

This research is supported by the Carmel Science Foundation. I would
like to thank Yael Oren, Arbel M. Ongo, and Ayelet B. Lata for
helpful discussions.


\begin{thebibliography}{99}
\bibitem{Noteunit} We use natural units in which $G=c=\hbar=1$.
\bibitem{Notepa} Here $M, Ma, Q$,
and $r_+=M\pm(M^2-a^2-Q^2)^{1/2}$ are respectively the mass, angular
momentum, charge, and horizon-radii of the black hole.
\bibitem{Delta} The parameter $\delta$ is closely related to the
angular-eigenvalue $\lambda$ of the angular Teukolsky equation, see
\cite{Teuk} for details [see, in particular, equations (2.7) and
(6.3) of \cite{Teuk}].
\end{thebibliography}
\end{document}